\documentclass[%
 reprint,doublecolumn, notitlepage,
 amsmath,amssymb,
 aps,
]{revtex4-2}

\usepackage{graphicx}
\usepackage{dcolumn}
\usepackage{bm}
\usepackage{lipsum}
\usepackage{graphicx}
\usepackage{hyperref}

\usepackage{tikz}
\hypersetup{%
        colorlinks=true,
        linkcolor=blue,
        citecolor=blue,
        urlcolor=blue,
      }

\usepackage[T1]{fontenc}
\usepackage[pass]{geometry}
\usepackage{lipsum}
\begin{document}


\title{Floating tunable coupler for scalable quantum computing architectures}

\author{Eyob A. Sete, Angela Q. Chen, Riccardo Manenti, Shobhan  Kulshreshtha, and Stefano Poletto}

\affiliation{%
Rigetti Computing,775 Heinz Avenue, Berkeley, CA 94710
}%

\date{\today}

\begin{abstract}
We propose a floating tunable coupler that does not rely on direct qubit-qubit coupling capacitances to achieve the zero-coupling condition. We show that the polarity of the qubit-coupler couplings can be engineered to offset the otherwise constant qubit-qubit coupling and attain the zero-coupling condition when the coupler frequency is above or below the qubit frequencies. We experimentally demonstrate these two operating regimes of the tunable coupler by implementing symmetric and asymmetric configurations of the superconducting pads of the coupler with respect to the qubits. Such a floating tunable coupler provides flexibility in designing large scale quantum processors while reducing the always-on residual couplings.

\end{abstract}

\pacs{Valid PACS appear here}
\maketitle
\section{Introduction}
The Realization of high-fidelity entangling gates for fault-tolerant quantum computation requires the addressing of competing sources of error. In particular, when scaling the quantum processor size, an always-on $ZZ$ coupling and an effective qubit-qubit coupling are two sources of coherent error that limit the performance of two-qubit gates in superconducting architectures. A scheme based on tunable couplers~\cite{Oliver18} has recently been introduced as a leading candidate for eliminating the always-on qubit-qubit coupling during the idling period between fast entangling gate operations.
The coupling element in this architecture consists of a grounded tunable coupler, i.e., it has a single superconducting pad connected to ground via a superconducting  quantum interference  device (SQUID). A virtual exchange interaction mediated by the tunable coupler can offset the direct qubit-qubit coupling. The zero-coupling condition is guaranteed when the frequency of the tunable coupler is above the frequency of the qubits and when there is a finite direct qubit-qubit coupling capacitance. Grounded tunable couplers have been explored as coupling elements between superconducting qubit pairs~\cite{Oliver18,Martinis19,Dapeng20,Wallraff20,Foxen20,Li20,Xu20,Oliver20}. Earlier studies have used inductive coupling~\cite{Geller15,Chen14} and a tunable bus~\cite{Tsai06,McKay16,Mundada19} as a coupling element. Recently, a tunable-coupler architecture connecting fixed-frequency floating qubits has been introduced~\cite{Stehlik21} in which a zero-coupling condition is attained when the tunable-coupler frequency is either above or below the qubit frequencies, and with a direct qubit-qubit coupling capacitance.

Thus far, all implementations of tunable couplers rely on a non-negligible direct coupling capacitance between the qubits, imposing restrictions on the circuit design because the qubits cannot be placed too far apart. In this work, we propose a \emph{floating} tunable coupler---in which two superconducting pads are connected by a SQUID---to achieve a vanishing net qubit-qubit coupling \emph{without} the need for a direct capacitive coupling between the qubits. This allows for a longer qubit-qubit pitch, which opens up space for readout resonators, Purcell filters, through-silicon vias, or other forms of three-dimensional integration required to scale up the size of the quantum processor. 

\begin{figure}
    \centering
    \includegraphics{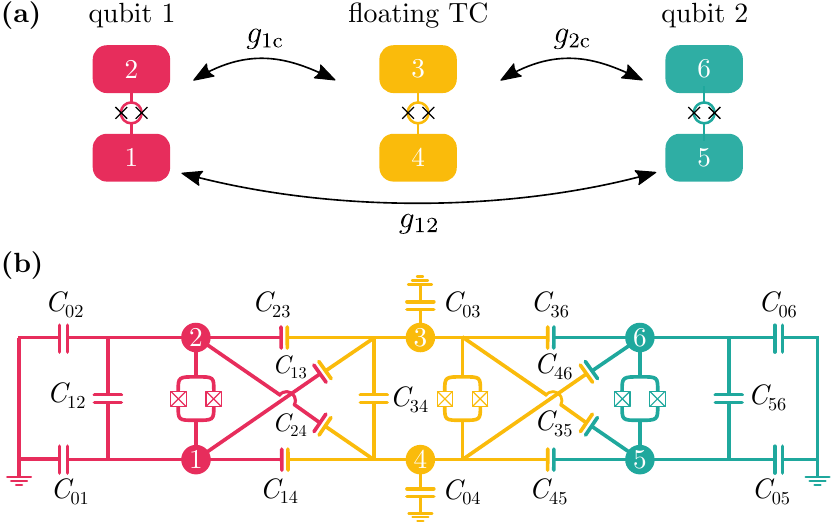}
    \caption{(a) A schematic of a floating tunable coupler (yellow) mediating the coupling between two floating qubits (fuchsia and teal). The qubit-coupler coupling is described by $g_{1c},g_{2c}$, and the direct qubit-qubit coupling is described by $g_{12}$. By tuning the frequency of the coupler, the effective coupling mediated by the coupler can offset the direct coupling $g_{12}$. (b) A lumped-element circuit representation of the schematic in (a), where the flux nodes represent the electrodes in the schematic. As discussed in the text, direct capacitance between the qubits is not needed to achieve the zero-coupling condition.
    }
    \label{fig:general_case}
\end{figure}

We introduce two operating regimes of the tunable coupler depending on the configurations of the superconducting pads of the coupler with respect to the qubits. In a symmetric configuration, where the qubits are strongly coupled to different pads of the coupler, the zero-coupling condition is achieved when the coupler frequency is below the qubit frequencies. On the contrary, in an asymmetric configuration, where the qubits are strongly coupled to the same pad of the coupler, the zero-coupling condition is attained when the coupler frequency is above the qubit frequencies. By designing devices with symmetric and asymmetric coupler configurations, we experimentally demonstrate the two operating regimes. This flexibility holds for both floating and grounded qubits, implying that even grounded qubits can reap the benefits of reaching the zero-coupling point when the coupler frequency is below the qubit frequencies.

\section{Theory of the tunable coupler}
We first consider the general case of two transmon qubits coupled via a tunable coupler as illustrated in Fig.~\ref{fig:general_case}. The qubit-coupler-qubit system can be described by the Hamiltonian
\begin{align}
H & = \sum_{k\in\{1,2,c\}}\left[\omega_{k}+\frac{E_{Ck}}{2} -\frac{E_{Ck}}{2}a_k^{\dag}a_k\right]a_k^{\dag}a_k\notag\\
    &\phantom{=}+ {\sum_{j=1}^2g_{jc}\left(a_{j}a_{c}^{\dag}+a_{j}^{\dag} a_{c}-a_{j}a_{c}-a_{j}^{\dag} a_{c}^{\dag}\right)} \notag \\
    &\phantom{=}+ g_{12}\left(a_{1}a_{2}^{\dag}+a_{1}^{\dag} a_{2}-a_{1}a_{2}-a_{1}^{\dag} a_{2}^{\dag}\right),
\end{align}
where $\omega_{k}/2\pi$, $E_{Ck}$, and $E_{Jk}$ are the frequencies, charging energies, and Josephson energies for the qubits ($k=1,2$) and tunable coupler ($k=c$), respectively, and $a_k$ $ (a_k^{\dag})$ is the annihilation (creation) operator. The qubit-coupler and qubit-qubit couplings are
\begin{align}
    g_{jc} &= \frac{E_{jc}}{\sqrt{2}}\left(\frac{E_{Jj}}{ E_{Cj}} \frac{E_{Jc}}{E_{Cc}}\right)^{\frac{1}{4}},~~j\in\{1,2\},\label{virtual}\\
    g_{12} &= \frac{E_{12}}{\sqrt{2}}\left(\frac{E_{J1}}{ E_{C1}} \frac{E_{J2}}{E_{C2}}\right)^{\frac{1}{4}}. \label{g_direct}
\end{align}
Approximating the qubits and the tunable coupler by their first three energy levels, applying a second-order Schrieffer-Wolff transformation, and assuming that the coupler remains in the ground state $\langle \sigma_{zc}\rangle = -1$, one obtains the qubit-qubit Hamiltonian \cite{Sete21}
\begin{align}
    &H_{qq}  = \sum_{j=1}^2\omega_{01,j} |1\rangle_{j}\langle 1| + (\omega_{02,j}-\eta_j)|2\rangle_j \langle 2| \notag\\
    &+ (g|10\rangle\langle 01| + g_{02} |11\rangle \langle 02|+ g_{20} |11\rangle \langle 20| +\mathrm{h.c.}),
\end{align}
where  $\omega_{01,j} = \omega_j  + \frac{g_{jc}^2}{\Delta_j} + \frac{g_{jc}^2}{\Sigma_j}$ and 
$\omega_{02,{j}} = 2\omega_j  + \frac{2g_{jc}^2}{\Delta_j+\eta_j} + \frac{2g_{jc}^2}{\Sigma_j-\eta_j}$ are the dressed eigenfrequencies of the qubits, and $\eta_j$ is the absolute value of the anharmonicity of the qubits. Here,  $g_{02}$ and $g_{20}$ represent the couplings between $|11\rangle$ and $|02\rangle$ and between $|11\rangle$ and $|20\rangle$, respectively. The net qubit-qubit coupling is 
\begin{align}
    g &= g_{12} -g_{\rm eff}, ~~ g_{\rm eff} = \frac{g_{1c}g_{2c}}{2}\sum_{j=1}^2\left(\frac{1}{\Delta_j} + \frac{1}{\Sigma_j}\right),\label{gqq}
\end{align}
where $\Delta_j = \omega_c-\omega_j$, $\Sigma_j = \omega_c+\omega_j$, and $g$ is the coupling strength between $|01\rangle$ and $|10\rangle$.

In order to achieve a vanishing qubit-qubit coupling $g=0$, the effective coupling $g_{\rm eff}$ that originates from a virtual exchange interaction through the coupler should compensate the direct qubit-qubit coupling $g_{12}$ at a certain coupler frequency. In general, there are two operating regimes that can satisfy this condition.
When $g_{12}$ and $g_{1c}g_{2c}$ have the same sign, the coupler frequency needs to be above the qubit frequencies ($\Delta_j>0$) to guarantee $g=0$. Alternatively, when $g_{12}$ and $g_{1c}g_{2c}$ have different signs, $g=0$ can be reached when $\Delta_j<0$, which means that the coupler frequency needs to be lower than the qubit frequencies. These two operating regimes are determined by the arrangement of the qubits and  coupler capacitor pads, as discussed in the next section. 

We note that most of the previous demonstrations of the tunable coupler utilize grounded qubits coupled to a grounded tunable coupler~\cite{Oliver18,Dapeng20,Wallraff20,Li20,Dapeng20,Xu20}.
In these cases, the zero-coupling condition occurs when the coupler frequency is above the qubit frequencies because $g_{12}$ and $g_{1c}g_{2c}$ have the same sign. In the recent demonstration in Ref.~\cite{Stehlik21}, the new zero-coupling regime ($g=0$ when coupler frequency is below the qubit frequencies) is accomplished by flipping the sign of $g_{12}$ through the use of floating qubits.
Both approaches make use of a direct qubit-qubit capacitive coupling to achieve the zero-coupling condition.

In the following, we consider a floating tunable coupler and show that it is possible to achieve $g=0$ \emph{without} a direct capacitance between the qubits.
Moreover, we show that the zero-coupling condition can be reached with the tunable-coupler frequency above or below the qubit frequencies for both floating and grounded qubits.

\begin{figure}
    \centering
    \includegraphics[width=\linewidth]{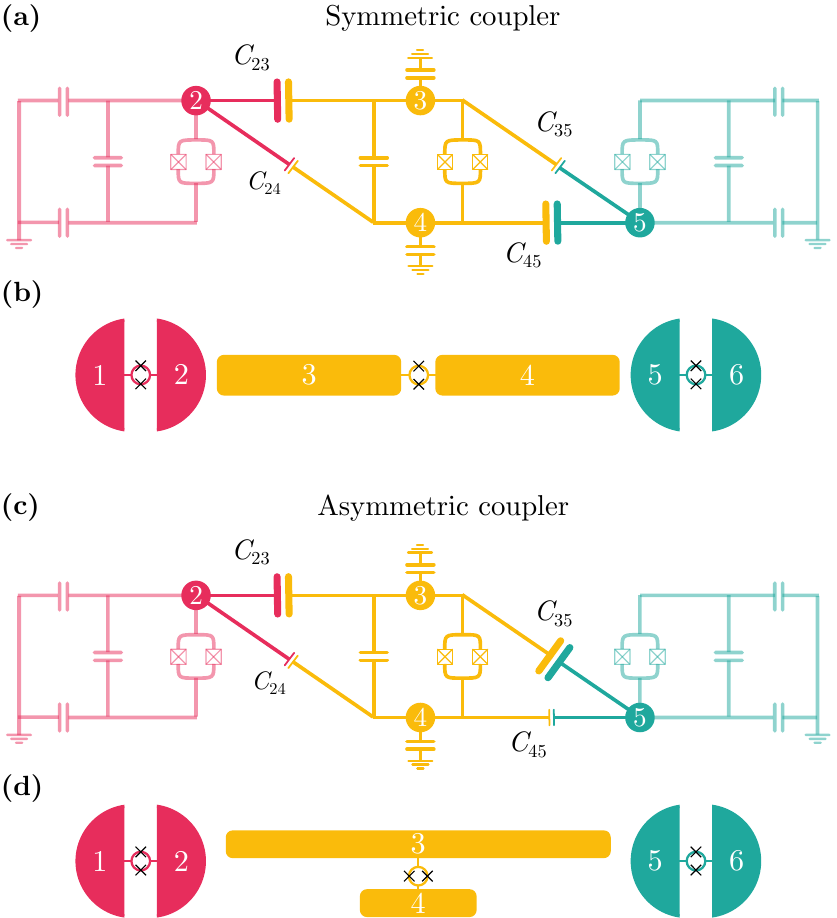}
    \caption{
    \textbf{Symmetric and asymmetric couplers}.
    (a) The lumped-circuit model for the symmetric coupler configuration with an example schematic (b). The coupling elements are highlighted, leaving the rest of the circuit washed out. The relative values of the coupling capacitances are represented by the size of the lumped elements (not to scale). The two largest coupling capacitances are connected to two different pads of the tunable coupler. (c) The lumped-circuit model for the asymmetric coupler configuration with an example schematic (d). The two largest coupling capacitances are connected to the same pad of the tunable coupler.
    The zero-coupling condition is reached when the tunable-coupler frequency is below (above) the qubit frequencies for the symmetric (asymmetric) configuration.
    }
    \label{fig:floating}
\end{figure}

\section{Floating tunable coupler with floating qubits}\label{floating qubits}
We consider a generic schematic of two floating tunable transmon qubits coupled via a floating tunable coupler as illustrated in Fig.~\ref{fig:general_case}(a). The general Hamiltonian describing the full system is derived in Appendix~\ref{floating}. The full lumped-element circuit model is shown in Fig.~\ref{fig:general_case}(b), where the ground node is represented by 0 and the other six nodes represent the electrodes of the qubits and the coupler. The capacitance variables are numbered based on how the nodes are connected, so that $C_{i,i+1}$ represents the capacitance between two neighboring electrodes. The qubits and tunable-coupler capacitances $C_{12}$, $C_{34}$, and $C_{56}$ include the parallel Josephson-junction capacitances. 

Following the standard quantization procedure \cite{Devoret97}, the Hamiltonian of the system can be written as 
\begin{align}\label{H_floating}
    H & =4E_{C1}\hat n_1^2+4E_{C2}\hat n_2^2 + 4E_{Cc}\hat n_c^2 \notag\\
    &+ 4E_{12}\hat n_1 \hat n_2 + 4E_{1c} \hat n_1 \hat n_c + 4E_{2c}\hat n_2 \hat n_c\notag\\
    &-\sum_{k\in \{1,2,c\}}E_{Jk}\cos(\hat\phi_{km}+\phi_{0k}),
\end{align}
where  $E_{Jk} (\phi_{ek}) = \sqrt{E_{JSk}^2+E_{JLk}^2+2E_{JSk}E_{JLk}\cos(\phi_{ek})}$,
$\phi_{0k} = \tan^{-1}\left[\frac{E_{JSk}-E_{JLk}}{E_{JSk}+E_{JLk}} \tan(\phi_{ek}/2)\right]\label{phi0}$; $n_{k} = Q_{k}/2e$, $E_{Ck}$ ($k\in \{1,2,c\}$) are Cooper-pair number operators and charging energies, respectively; $E_{12}$, $E_{1c}$, and $E_{2c}$ are coupling energies, and $E_{JLk}$ and $E_{JSk}$ are the Josephson energies of the two junctions of the SQUID ; $\Phi_{k}$ are the node fluxes, $\phi_{ek} = 2\pi \Phi_{ek}/\Phi_0$ being the reduced external flux biases through the SQUID loops. The reduced fluxes are defined as $\phi_{l} = 2\pi \Phi_{l}/\Phi_{0}$ ($l\in \{1,2,...6\}$) and $\hat\phi_{1m} = \hat\phi_2-\hat\phi_1$, $\hat \phi_{2m} = \hat \phi_6-\hat\phi_5$, $\hat\phi_{cm} = \hat\phi_4-\hat\phi_3$ with $\Phi_0= h/2e$ being the flux quantum.

\begin{figure}
    \centering
    \includegraphics{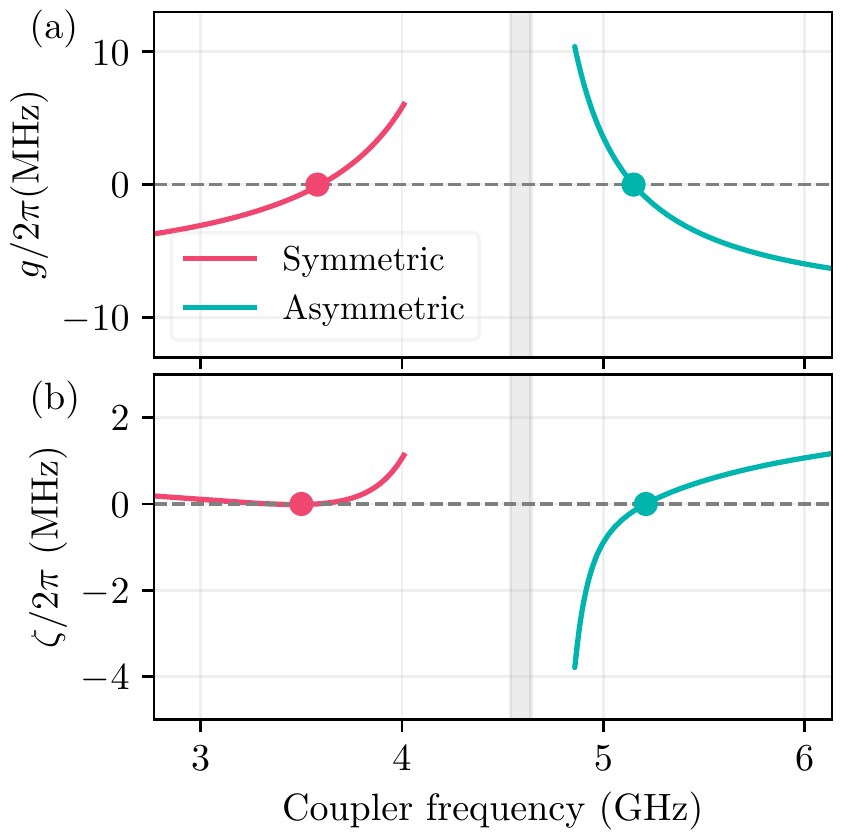}
    \caption{\textbf{Qubit-qubit couplings and residual $ZZ$ using a floating coupler}. (a) The simulated net qubit-qubit coupling and (b) the residual $ZZ$ interaction for symmetric (fuchsia) and asymmetric (teal) configurations when qubits are parked at the dc sweet spot versus tunable-coupler frequency for realistic design parameters (see main text for details).
    The gray area highlights the maximum frequency band of the qubits.
    The two different regimes of zero coupling can be engineered by selecting the desired layout of the coupler.
    }
    \label{fig:g_vs_coupler_floating_qubits}
\end{figure}

For simplicity, we assume that all capacitances from the qubit pads to ground are the same: $C_g=C_{01}=C_{02}=C_{05}=C_{06}$. We make the same assumption for the tunable-coupler pads, $C_{gc}=C_{03}=C_{04}$.
The qubits and tunable-coupler capacitances, which include the parallel Josephson-junction capacitances, are denoted by $C_q=C_{12}=C_{56}$ and $C_c=C_{34}$. We also make the experimental assumption that the floating tunable coupler is coupled to only one superconducting pad on each qubit and that all qubit-coupler coupling capacitances $C_{i,j}$ are smaller than the qubit capacitances and capacitances to the ground, i.e., $C_{i,j}\ll \{C_{c},C_{g},C_{gc},C_q\}$.
With these assumptions, the coupling energies are approximately given by
\begin{eqnarray}
    E_{12}  &\approx & -\frac{e^2}{\tilde C} 
    \big\{[(C_{23}+C_{24})(C_{45}+C_{35})+C_{24}C_{35}]C_c \notag\\
    &&+(C_{23}C_{35}+C_{45}C_{24})C_{gc}\big\},\label{e12}\\
    E_{1c} &\approx& -\frac{e^2(C_{23} - C_{24})}{(2 C_q + C_g) (2 C_c + C_{gc})},\label{Ec1}\\
    E_{2c} &\approx& -\frac{e^2(C_{45} - C_{35})}{(2 C_q + C_g) (2 C_c + C_{gc})},\label{Ec2}
\end{eqnarray}
where $\tilde C = C_{gc}(2 C_q + C_g)^2 (2 C_c + C_{gc})$. It is worth mentioning that in these equations, we assume that the qubits are sufficiently far apart so that the direct qubit-qubit capacitances are negligible. As a result, the derived coupling energies do not depend on the direct qubit-qubit capacitance. 

Although the magnitudes of $E_{12}$, $E_{1c}$, and $E_{2c}$ are determined by the designed capacitances, we see from Eqs.~\eqref{e12}-\eqref{Ec2} that $E_{12}$ is always negative and that the sign of $E_{1c}$ ($E_{2c}$) depends on the relative magnitudes of the nearest-neighbor capacitance $C_{23}$ ($C_{45}$) and the next-nearest-neighbor capacitance $C_{24}$ ($C_{35}$). Assuming that the magnitude of the nearest-neighbor and next-nearest-neighbor qubit-coupler capacitances are not the same, we identify two distinct design layouts, which we refer to as \emph{symmetric} and \emph{asymmetric} coupler configurations. These configurations will determine whether the coupler frequency has to be above or below the qubit frequencies to achieve the zero-coupling condition. 

In the symmetric configuration [Fig~\ref{fig:floating}(a)], the two largest coupling capacitances are connected to different pads of the tunable coupler. This can be realized, for example, by placing the tunable-coupler pads symmetrically between the qubits [Fig~\ref{fig:floating}(b)] such that the dominant capacitances $C_{23}$ and $C_{45}$ occur between the qubit pads and different tunable-coupler pads. In this case, we have the relation $\{C_{23}, C_{45}\} > \{C_{24}, C_{35}\}$. Therefore, in the symmetric configuration, it follows from Eqs.~\eqref{virtual}, \eqref{g_direct}, and  Eqs.~\eqref{e12}-\eqref{Ec2} that the couplings $g_{kc}$ and $g_{12}$ are all negative. To achieve the zero-coupling condition, the tunable-coupler frequency has to be below the qubit frequencies [Eq.~\eqref{gqq}] so that $g_{\rm eff}$ can offset the fixed direct coupling $g_{12}$. 

If the largest coupling capacitances are connecting the qubits to the same tunable-coupler pad, we have the asymmetric configuration [Fig~\ref{fig:floating}(c)]. An example of an asymmetric tunable-coupler configuration is shown in Fig.~\ref{fig:floating}(d). In this asymmetric configuration, we have $\{C_{23}, C_{35}\} > \{C_{24}, C_{45}\}$. Thus, the fixed coupling $g_{12}$ is negative just as in the symmetric configuration, but this time the couplings $g_{1c}$ and $g_{2c}$ have opposite signs [see Eqs.~\eqref{Ec1} and \eqref{Ec2}].
As a result, the zero-coupling condition is achieved with the tunable-coupler frequency above the qubit frequencies as per Eq.~\eqref{gqq}. 

\begin{figure*}[t]
    \centering
    \includegraphics[width=\linewidth]{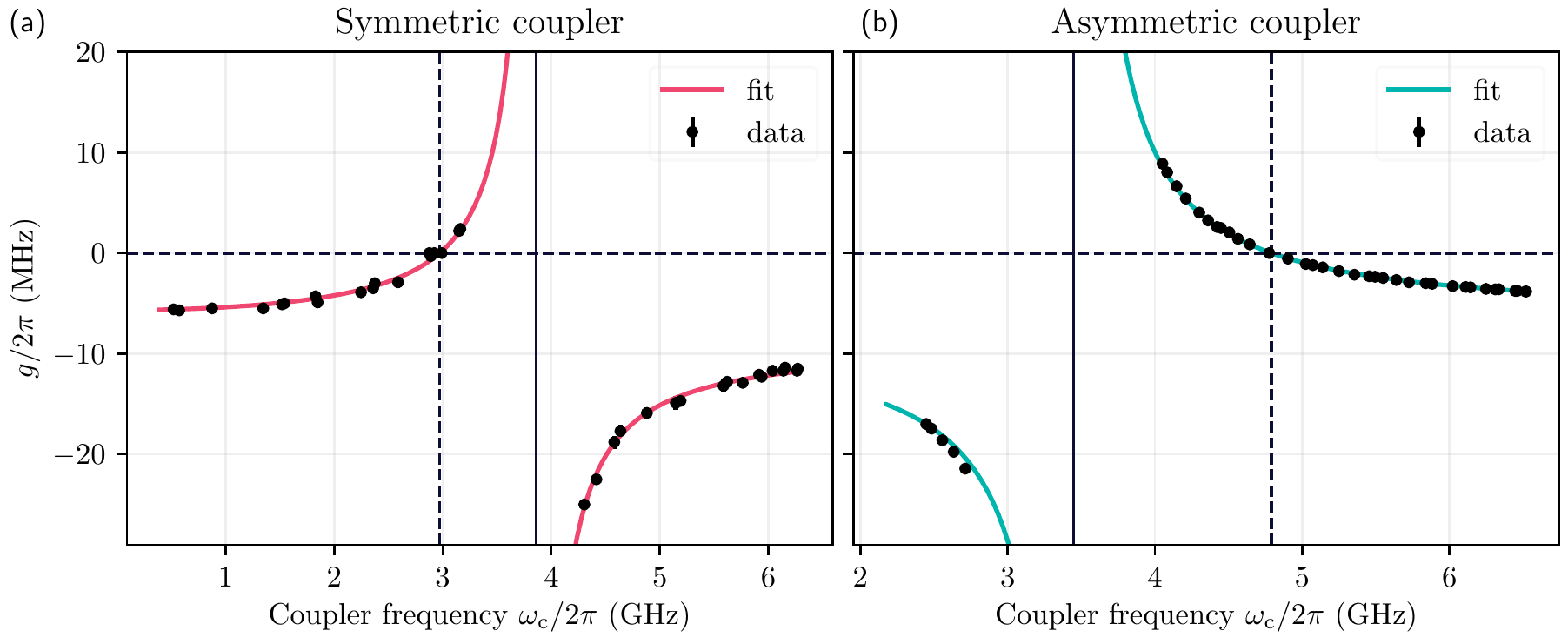}
    \caption{{{\bf Zero-coupling operating regimes on measured devices}}. 
    The net qubit-qubit coupling $g$ versus the frequency of the tunable coupler. (a) The zero-coupling point occurs when $\omega_c < \{\omega_{1},\omega_{2}\}$ for the symmetric coupler since $g_{12}$ and $g_{1c}g_{2c}$ have different signs. (b) For the asymmetric coupler configuration, the zero-coupling point occurs when $\omega_c>\{\omega_{1},\omega_{2}\}$ since $g_{12}$ and $g_{1c}g_{2c}$ now have the same sign. The vertical dashed lines show the coupler frequency at which the coupling is zero, while the vertical solid line indicates the resonance condition of the two qubits. In these plots, the couplings are measured for a full period and the coupler frequency is obtained from the fits. For the symmetric (asymmetric) coupler, we use level-splitting (parametric-chevron) measurement to extract $g$.}
    \label{fig:data}
\end{figure*}

As an example,  we numerically calculate the qubit-qubit coupling for qubit-coupler-qubit system capacitances from electromagnetic simulations: $C_{g} = 110~\mathrm{fF}$, $C_q = 46~\mathrm{fF}$, $C_{gc} = 80~\mathrm{fF}$, $C_{c} = 61~\mathrm{fF}$, $C_{24}=2~\mathrm{fF}$, $C_{23}=19.5~\mathrm{fF}$ (other coupling capacitances are less than 0.4 fF and their contributions to the net qubit-qubit coupling are negligible). The qubit frequencies are assumed to be in the bands $f_1 = 4.1-4.58$ GHz and $f_2 = 4.1-4.64$ GHz for both symmetric and asymmetric couplers. For the asymmetric coupler, we use $C_{35}=19.5~\mathrm{fF}$, $C_{45}=2~\mathrm{fF}$, and a coupler frequency above the qubit frequencies, $f_c= 4.8-6.14~\mathrm{GHz}$. The corresponding couplings are $g_{12}/2\pi=-12~\mathrm{MHz}$, $g_{1c}/2\pi = -79~\mathrm{MHz}$, and $g_{2c}/2\pi=98~\mathrm{MHz}$. For the symmetric case, we use $C_{35}=2~\mathrm{fF}, C_{45}=19.5~\mathrm{fF}$, and a coupler frequency $f_c = 2.77-4.0~\mathrm{GHz}$ , yielding coupling rates $g_{12}/2\pi=-5.8~\mathrm{MHz}$, $g_{1c}/2\pi = -85~\mathrm{MHz}$, and $g_{2c}/2\pi=-85~\mathrm{MHz}$. With these parameters, the $g=0$ condition is reached for frequencies of the tunable coupler below (above) the qubit frequencies in the symmetric (asymmetric) case [Fig.~\ref{fig:g_vs_coupler_floating_qubits}(a)]. We also numerically compute the residual $ZZ$ interaction strength  $\zeta = \omega_{\overline{|11\rangle}}-\omega_{\overline{|10\rangle}}-\omega_{\overline{|01\rangle}}+\omega_{\overline{|00\rangle}}$ ($\omega_{\overline{j}}$ are the eigenfrequencies of the coupled qubits) as a function of the coupler frequency with the qubits parked at their maximum frequency [Fig.~\ref{fig:g_vs_coupler_floating_qubits}(b)]. The residual $ZZ$ coupling is zero at a slightly different flux bias than $g=0$ because of repulsion from higher energy levels.

\section{Experimental validation}
We experimentally verify the two distinct operating regimes by measuring two different devices. One is designed with a layout as shown in Fig.~\ref{fig:floating}(b), while the second is designed as shown in Fig.~\ref{fig:floating}(d). The measured devices consist of two floating tunable transmon qubits, each capacitively coupled to a floating tunable coupler. The tunable coupler is designed with a symmetric SQUID and a maximum frequency above the qubit frequencies so that the coupler frequency can be parked above or below the qubit frequencies during the experiments. The experimental setup is discussed in Appendix \ref{Device_parameters}.

The magnitude of the net qubit-qubit coupling $g$ can be extracted either from the energy-level splitting in qubit spectroscopy or from chevron measurements, which are performed at different tunable-coupler dc flux biases. An alternative way to characterize the net qubit-qubit coupling is by measuring the energy exchange rates between the two qubits using parametric-chevron measurements \cite{Sete21}. This can be done by first applying a $\pi$ pulse on one of the qubits and bringing the qubits into resonance, which is done at different coupler dc flux biases. Since the spectroscopic and chevron measurements only provide the magnitude of $g$, negative signs are added based on our knowledge of the tunable-coupler system, allowing for easier comparison with theory. By fitting the measured $g$ versus flux data to Eq.~\eqref{gqq}, we can extract $g_{12}$ and the product $g_{1c}g_{2c}$ as well as the tunable-coupler frequency $\omega_c$. The qubit and coupler parameters from the fit are shown in Appendix \ref{Device_parameters}.

We first look at a device with a symmetric coupler, the design of which is shown in Fig.~\ref{fig:floating}(b). The measured $g$ for the symmetric coupler is plotted in Fig.~\ref{fig:data}(a). In agreement with theory, the net coupling never crosses $g=0$ when the coupler frequency is above the qubit frequencies. When the coupler frequency is below the qubit frequencies, the virtual coupling $g_{\rm eff}$ becomes negative so that $|g_{12}|= |g_{\rm eff}|$, satisfying the condition $g=0$ (illustrated by the dashed vertical lines). At the $g=0$ point $\omega_c/2\pi = 2.97~\mathrm{GHz}$ and $g_{jc}/\Delta_j = 0.12$, indicating that the qubit-coupler system is in the dispersive regime. When operating entangling gates, the coupler can be tuned further down in frequency to minimize the dynamical $ZZ$ coupling  [Fig.~\ref{fig:g_vs_coupler_floating_qubits}(b)].

\begin{figure}[b]
    \centering
    \includegraphics[width=\linewidth]{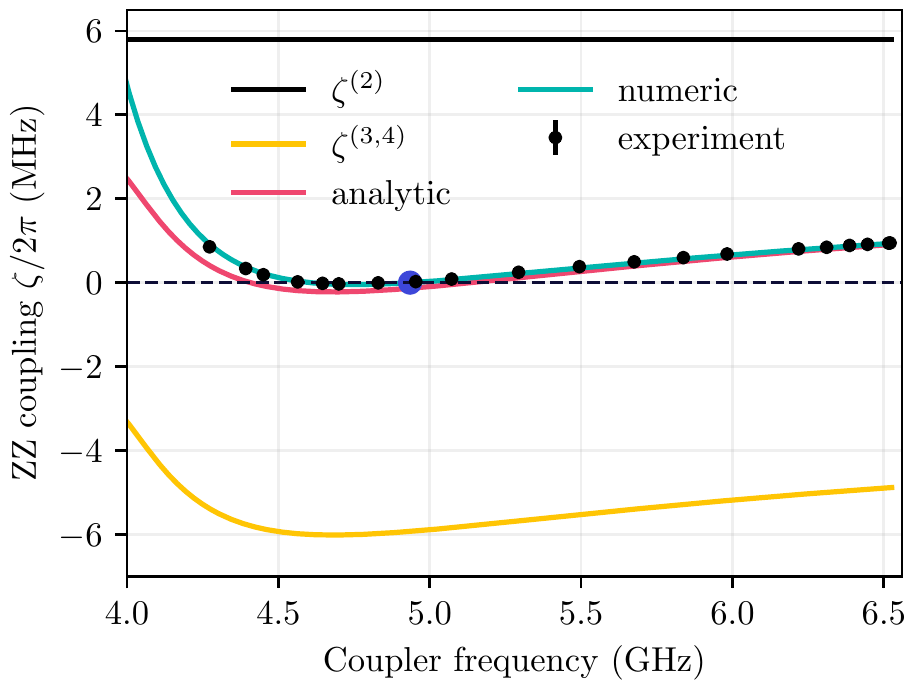}
    \caption{{\textbf{ Zero-$ZZ$ coupling on asymmetric tunable-coupler device}}.  The residual $ZZ$ coupling $\zeta$ versus the coupler frequency. The dotted points are measured values and the solid teal line represents the $ZZ$ values calculated numerically by diagonalizing the corresponding Hamiltonian constructed with measured parameters. The magenta curve represents the approximate analytical expression for $ZZ$ coupling. The black (yellow) line corresponds to the second (third plus fourth) term in the $ZZ$ expansion.}
    \label{fig:ZZ_experiment}
\end{figure}

We next consider the asymmetric coupler configuration shown in Fig.~\ref{fig:floating}(d). In this case, we implement the chevron measurement to characterize the coupling. The virtual coupling $g_{\rm eff}$ becomes negative when the tunable-coupler frequency is above the qubit frequencies, offsetting the fixed coupling $g_{12}$ at $\omega_c/2\pi = 4.79~\mathrm{GHz}$ [Fig.~\ref{fig:data}(b)]. The qubit-coupler system is in the dispersive regime at the $g=0$ flux bias, $g_{jc}/\Delta_j = 0.1$. We also measure the residual $ZZ$ coupling between the qubits for the asymmetric coupler by parking the qubits at their maximum frequency and varying the flux bias on the coupler. The $ZZ$ coupling is measured by measuring the frequency of one of the qubits using a Ramsey measurement while the other qubit is either in the ground or excited state. The difference in frequency from the two Ramsey measurements gives the $ZZ$ coupling. In Fig.~\ref{fig:ZZ_experiment}, we plot measured $ZZ$ as well as simulated (using measured parameters) $ZZ$ coupling as a function of the tunable-coupler frequency. The experiment agrees well with the numerical simulation. The residual $ZZ$ vanishes at $f_c=4.935~\mathrm{MHz}$. To gain insight into the dependence of $ZZ$ coupling on the coupler flux bias (or coupler frequency), we consider an approximate expression of the $ZZ$ coupling derived using perturbation expansion \cite{Li20}. The second-order expansion of the Hamiltonian with respect to coupling yields the leading term in $ZZ$ coupling
\begin{align}
\zeta^{(2)}=-\frac{2g_{12}^2(\eta_1+\eta_2)}{(\Delta_{12}-\eta_1)(\Delta_{12}+\eta_2)},
\end{align}
where $\Delta_{12}=\omega_1-\omega_2$ and the $\eta_j$ are anharmonicities. The flux-dependent contribution to $ZZ$ coupling can be accounted for by keeping the third- and fourth-order terms in the expansion
\begin{align}
    \zeta^{(3,4)}(\Phi_{ec})&= -\frac{2g_{12}g_{1c}g_{2c}}{\Upsilon^2(\Phi_{ec})}\Big[
    \frac{1}{\Delta_{2}}\left(\frac{1}{\Delta_{12}}+\frac{2}{-\Delta_{12}+\eta_1}\right)\notag\\
    &+\frac{1}{\Delta_{1}}\left(\frac{2}{\Delta_{12}+\eta_2}-\frac{1}{\Delta_{12}}\right)
    \Big]\notag\\
    &-\frac{2g_{1}^2g_{2c}^2}{(\Delta_{1}+\Delta_{2}+\eta_c)\Upsilon^4(\Phi_{ec})}\left(\frac{1}{\Delta_{1}}+\frac{1}{\Delta_{2}}\right)^2\notag\\
    &+\frac{g_{1c}^2g_{2}^2}{\Delta_{1}^2\Upsilon^4(\Phi_{ec})}\left(\frac{2}{\Delta_{12}+\eta_2}-\frac{1}{\Delta_{12}}+\frac{1}{\Delta_{2}}\right)\notag\\
    &+\frac{g_{1c}^2g_{2c}^2}{\Delta_{2}^2\Upsilon^4(\Phi_{ec})}\left(\frac{2}{-\Delta_{12}+\eta_1}+\frac{1}{\Delta_{12}}+\frac{1}{\Delta_{1}}\right),
\end{align}
where the term $\Upsilon(\Phi_{ec})=[E_{Jc}(0)/E_{Jc}(\Phi_{ec})]^{1/4}$ accounts for the flux-dependent qubit-coupler couplings. The approximate analytical expression for residual $ZZ$ is then $\zeta (\Phi_{ec})\approx \zeta^{(2)}+\zeta^{(3,4)}(\Phi_{ec})$.

For the asymmetric coupler device, we measure $\Delta_{12}/2\pi = 182~\mathrm{MHz}$, $\eta_{1}/2\pi=219~\mathrm{MHz}$, $\eta_{2}/2\pi=215~\mathrm{MHz}$, and $g_{12}/2\pi=-9.4~\mathrm{MHz}$. Therefore, the leading ZZ term $\zeta^{(2)}$ is positive and constant, ($\zeta^{(2)}/2\pi = 5.55~\mathrm{MHz}$; see the black line in  Fig.~\ref{fig:ZZ_experiment}). On the contrary, the $ZZ$ coupling $\zeta^{(3,4)}$ that includes third- and forth-order terms depends on the coupler flux bias and is negative (the yellow curve in Fig.~\ref{fig:ZZ_experiment}). As the frequency of the coupler decreases,  $\zeta^{(3,4)}$ increases in magnitude and fully compensates the contribution from the static coupling, yielding the first $\zeta=0$ point (the blue dot in Fig.~\ref{fig:ZZ_experiment}). The magnitude of $\zeta^{(3,4)}$ become slightly larger than $\zeta^{(2)}$ before it rapidly decreases, giving the second $\zeta=0$ point. The flux-dependent $ZZ$ coupling decreases in magnitude when the coupler frequency is decreased further, resulting in a larger net $ZZ$ coupling. As the coupler approaches the qubit frequencies, the fourth-order perturbation deviates from the experimental and numerical data.

\begin{figure}[t]
    \centering
\includegraphics{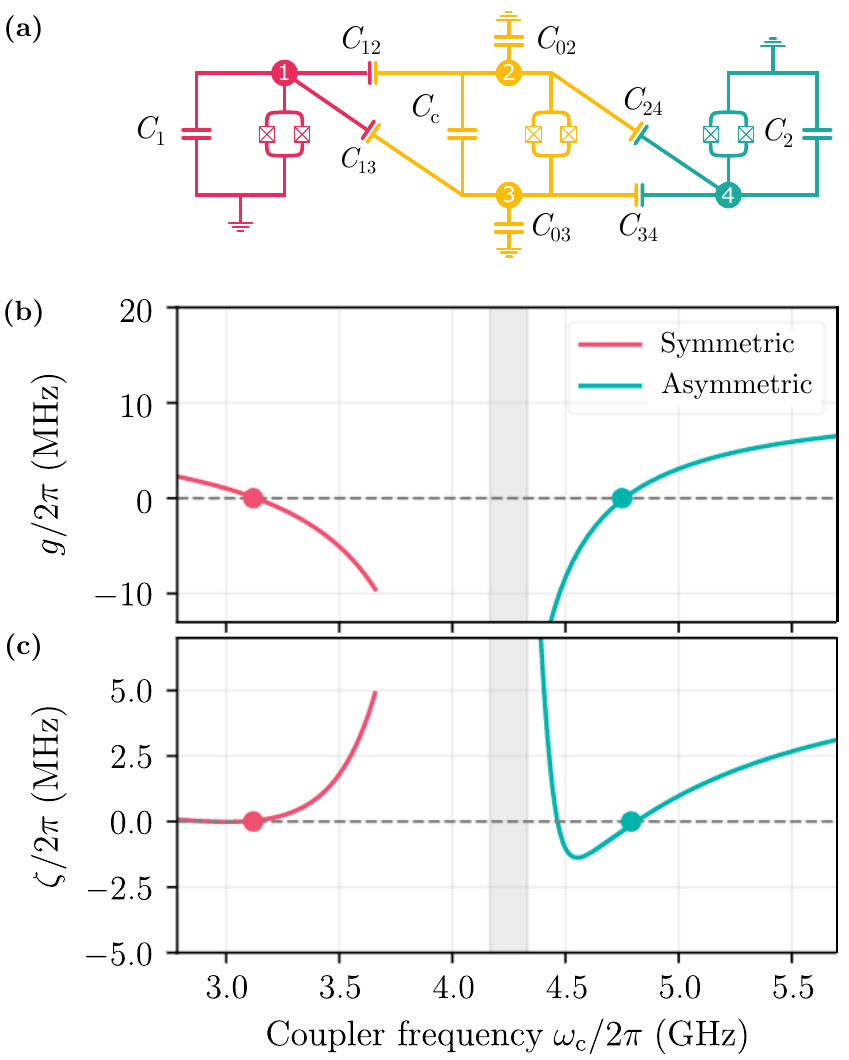}
    \caption{\textbf{Grounded qubits coupled via a floating coupler}. (a) A lumped-element circuit model of grounded qubits coupled via a floating tunable coupler. Similar to floating qubits coupled by a floating coupler, the polarity of $g_{jc}$ can be controlled by engineering the coupling capacitances $C_{12},C_{13},C_{24}$, and $C_{35}$ and thus offsetting the direct coupling $g_{12}$ by tuning the coupler above or below the qubit frequencies. (b) Simulated qubit-qubit couplings and (c) residual $ZZ$ interactions for symmetric (fuchsia) and asymmetric (teal) configurations when qubits are parked at the dc sweet spot versus tunable-coupler frequency for realistic design parameters (see main text for details). The gray area highlights the maximum frequency band of the qubits.}
    \label{fig:grounded}
\end{figure}

\section{Floating tunable coupler with grounded qubits}
Thus far, we have considered circuits where the floating tunable coupler is connected to floating qubits. However, the ability to engineer the zero-coupling operating regimes also holds for \emph{grounded} qubits connected to a floating coupler. Here, we describe the theory for this case but we do not present any experimental data.
 
Similar to Section \ref{floating qubits}, we can draw a lumped-element circuit model for a floating coupler-grounded qubit configuration [Fig.~\ref{fig:grounded}(a)]. The coupling energies that determine the signs of $g_{12}$ and $g_{kc}$ (see Appendix \ref{ground} for the full derivation) are given by 
\begin{align}
E_{12} & =\frac{e^{2}}{4C_{\text{tot}}}[C_{12}C_{34}(C_{13}+C_{24}+C_c)\nonumber\\
 & \qquad\qquad+C_{13}C_{24}(C_{12}+C_{24}+C_{c})\nonumber\\
 & \qquad\qquad+(C_{c}+C_{gc})(C_{12}C_{24}+C_{13}C_{34})],\\
E_{1c} & =-\frac{e^{2}}{C_{\text{tot}}}[C_{2}(C_{12}C_{34}-C_{13}C_{24})\nonumber\\
 & \qquad\qquad\;\;+C_{gc}C_{\Sigma2}(C_{12}-C_{13})],\\
E_{2c} & =\frac{e^{2}}{C_{\text{tot}}}[C_{1}(C_{12}C_{34}-{C_{13}}C_{24})\nonumber\\
 & \qquad\quad+C_{gc}C_{\Sigma1}(C_{34}-C_{24})],
\end{align}
where $C_{\Sigma 1(\Sigma 2)} = C_{1(2)}+C_{12(34)}+C_{13(24)}$ and we assume that the capacitances from the pads of the coupler to the ground are the same, $C_{gc}= C_{02}=C_{03}$. $C_{\rm tot}$ is the total capacitance and its explicit form is given in Appendix~\ref{ground}. Again, the coupling-energy equations do not depend on the direct qubit-qubit capacitance. 
The qubit-qubit coupling energy $E_{12}$ is always positive, while the qubit-coupler coupling energies $E_{1c}$ and $E_{2c}$ can be positive or negative depending on the relative magnitude of the coupling capacitances.

For the symmetric configuration,   $\{C_{12}, C_{34}\} > \{C_{13}, C_{24}\}$ and the coupling energy $E_{1c}$ becomes negative while $E_{2c}$ turns positive.
Since the signs of the coupling rates are determined by the signs of the coupling energies, we have $g_{12}>0$ and $g_{1c}g_{2c}<0$.
Therefore, according to Eq.~\eqref{gqq}, $g_{\rm eff}$ will become positive and can offset $g_{12}$ when the coupler frequency is below that of the qubits. A similar analysis for the asymmetric configuration, where $\{C_{12}, C_{24}\} > \{C_{13}, C_{34}\}$, leads to a zero-coupling condition for the tunable-coupler frequency above the qubit frequencies.

As an example, we consider realistic capacitances $C_{1} = 96~\mathrm{fF}$, $C_{2} = 97~\mathrm{fF}$, $C_{gc} = 110~\mathrm{fF}$, $C_{12}=10~\mathrm{fF}$, $C_{13}=1~\mathrm{fF}$ which are the same for symmetric and asymmetric configurations and qubit frequencies $f_1 = 4.18$ GHz and $f_2 = 4.54$ GHz. For the symmetric case, we consider a coupler below the qubit frequencies $f_{c} = 2.787-3.663~\mathrm{GHz}$ and capacitances $C_c = 52~\mathrm{fF}$, $C_{24}= 10~\mathrm{fF}$, and $C_{34}= 11~\mathrm{fF}$. Whereas for the asymmetric case, we set the coupler frequency to be above the qubit frequencies, $f_{c} = 4.38-5.71~\mathrm{GHz}$ and capacitances $C_c = 52~\mathrm{fF}$, $C_{24}= 1~\mathrm{fF}$, and $C_{34}=10~\mathrm{fF}$. In Figs.~\ref{fig:grounded}(b) and \ref{fig:grounded} (c),  we plot the simulated qubit-qubit coupling and the residual ZZ interaction versus the tunable-coupler frequency for both symmetric and asymmetric configurations. The net qubit-qubit coupling $g$ and the $ZZ$ coupling strength $\zeta$ are effectively zero at slightly different coupler frequency because of energy-level repulsion. The $ZZ$ coupling for the asymmetric coupler case has two zero-ZZ points similar to the results in Sev. IV. This is because the flux-dependent $ZZ$ term has opposite sign than the static $ZZ$ term and crosses it twice within the band of the coupler frequency, giving the two zero-$ZZ$ points. The coupler flux bias that gives $\zeta=0$ will be a parking flux bias (or idle point) where single qubit gates are operated. It is worth noting that vanishing $g$ and $\zeta$ are achieved without requiring a finite direct qubit-qubit capacitance. 

\section{Conclusion}
We propose and realize a floating tunable-coupler scheme that does not rely on a direct qubit-qubit capacitance to reach the zero-coupling condition. We show that we can engineer two operating regimes of the tunable coupler by varying the layout of the pads of the tunable coupler with respect to that of the qubits, which we refer to as symmetric and asymmetric configurations. In the symmetric configuration, the zero-coupling condition is achieved when the coupler frequency is below the qubit frequencies, whereas in the asymmetric configuration the coupler frequency has to be above the qubit frequencies to satisfy the zero-coupling condition. These operating regimes hold for both floating and grounded qubits when connected by a floating tunable coupler. A floating tunable coupler thus provides flexibility in designing a scalable superconducting qubit architecture that is immune to always-on residual interactions.

\begin{acknowledgements}
We thank Ani Nersisyan for helping with the general design process. We also thank Glenn Jones and Douglas Zorn for developing the control hardware used in the experiment.
\end{acknowledgements}

\section*{Contributions}

E.S. developed the theory of the floating tunable coupler for different operating regimes. E.S. and A.C. acquired the data and performed the data analysis. R.M. proposed the layout of the asymmetric coupler, simulated, and designed the devices. E.S., S.P., and A.C. wrote the manuscript. All authors contributed to the editing of the manuscript and figures. S.P. coordinated the effort.

\appendix
\begin{widetext}
\section{Experimental setup and device parameters}\label{Device_parameters}
The devices are mounted to a copper PCB using Al wirebonds, packaged in a light-tight assembly, and measured in a dilution refrigerator at 8~mK. The dc and microwave signals are delivered via non-magnetic sub miniature push-on (SMP) surface mount connectors. An overview of the experimental setup used to measure the device used in this experiment is shown in Fig.~\ref{fig:wiring}, where individual components are labelled.

The measured devices consist of two tunable transmons connected to a tunable coupler. The maximum frequency of the two qubits $f_{01}$, their anharmonicities $\eta$ as well as the extracted value for the maximum tunable coupler frequency for the devices measured in the main text are shown in Table~\ref{tab:qubit_parameters}. 

\begin{table}[h!]
    \centering
    \begin{tabular}{ccccccc}
    \hline
    \hline
    &\multicolumn{3}{c}{Symmetric}&
    \multicolumn{3}{c}{Asymmetric} \\
    \hline
      parameters   & q1 & qc &   q2 & q1 &qc& q2\\
        \hline
      
   $f_{01}$ (GHz)  &3.862 & 6.041  &4.045 & 3.449 &6.526 & 3.63\\
   $\eta/2\pi$(MHz)&230 &  - &233  & 219 & - & 215\\
    $T_{1}(\mu\mathrm{s})$  & 26& -&19& 38 &-& 10\\
    \hline
    \hline
     \end{tabular}
    \caption{Qubit parameters of the measured devices with symmetric and asymmetric couplers. The tunable coupler frequencies are extracted from a fit.}
    \label{tab:qubit_parameters}
\end{table}

\begin{figure}
    \centering
    \includegraphics{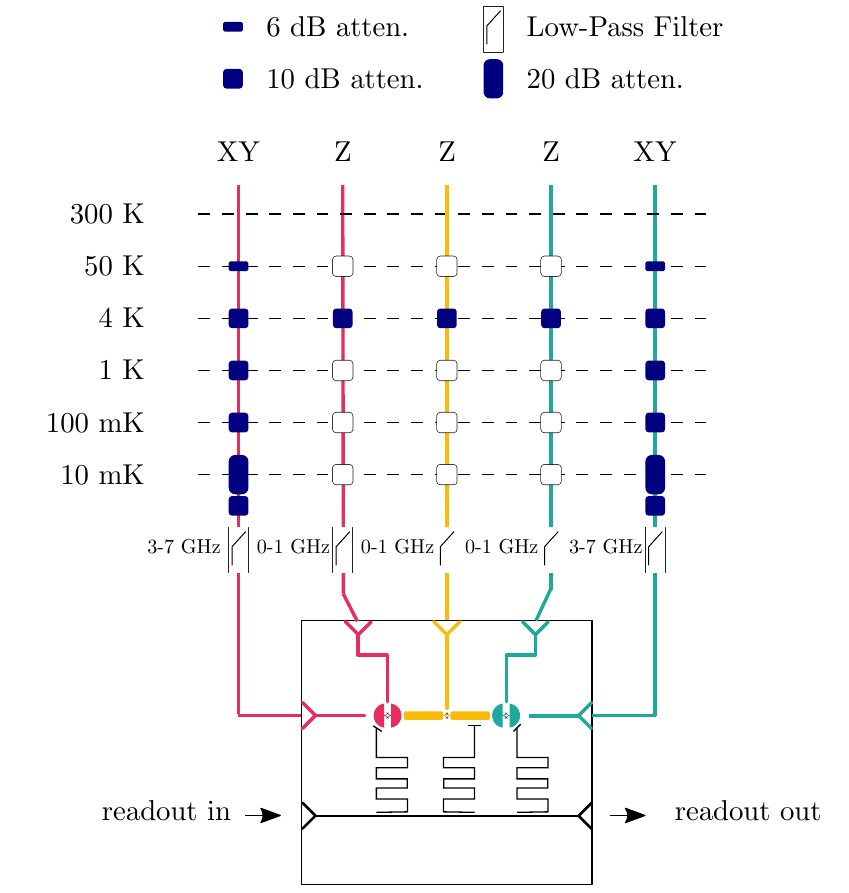}
    \caption{\textbf{The experimental setup}. This diagram details the control electronics, wiring, and filtering for
the two qubits and the coupler. The readout signal, rf signal (XY lines), and dc and ac signals for flux delivery (Z lines) are generated by custom Rigetti arbitrary waveform generators (AWG)s. All control lines go through various stages of attenuation and filtering in the dilution refrigerator.}
    \label{fig:wiring}
\end{figure}
The measured $g$ as a function of tunable coupler frequency (Fig.~\ref{fig:data}) was fit to Eq.~\ref{gqq}. In doing so, we  extracted the direct qubit-qubit coupling $g_{12}$ and the product $g_{1c}g_{2c}$ (Table~\ref{tab:coupling_parameters}). 
\begin{table}[h!]
    \centering
    \begin{tabular}{ccccc}
    \hline
    \hline
    &\multicolumn{2}{c}{Symmetric}&
    \multicolumn{2}{c}{Asymmetric} \\
    \hline
      parameters   & Design   & Measured & Design & Measured\\
        \hline
   $g_{12}/2\pi$ (MHz) & -9.1&-5.7& -14.2 & -9.4\\
   $\sqrt{g_{1c} g_{2c}}/2\pi$ (MHz) &111&107.8& 150 &131.6\\
    \hline
    \hline
    \end{tabular}
    \caption{Designed and measured coupling strengths for the devices with symmetric and asymmetric couplers. }
    \label{tab:coupling_parameters}
\end{table}

\section{Derivation of the Hamiltonian of floating qubits coupled via a floating tunable coupler}\label{floating}
Here we derive the Hamiltonian for two floating tunable qubits coupled via a floating tunable coupler. In the derivation, we include all eight coupling capacitances between the pads of the qubits and the tunable coupler, but we purposefully exclude the direct qubit-qubit capacitance. Node fluxes are numbered from 1 to 6 on the qubits and tunable coupler pads. Capacitances are numbered according to the nodes connected as $C_{mn}$, with $m<n$ (schematic shown in Fig.~\ref{fig:general_case}). The Hamiltonian of the lumped-element circuit model is derived starting from the Lagrangian of the circuit \cite{Devoret97,Girvin2014}
\begin{align}
\mathcal{L} &= T-U,\notag\\[1em]
T &= \frac{1}{2}C_{12}(\dot \Phi_{2}-\dot\Phi_1)^2+\frac{1}{2}C_{02}\dot\Phi_2^2+\frac{1}{2}C_{01}\dot\Phi_1^2+\frac{1}{2}C_{23}(\dot\Phi_2-\dot\Phi_3)^2+\frac{1}{2}C_{13}(\dot\Phi_1-\dot\Phi_3)^2+\frac{1}{2}C_{24}(\dot\Phi_2-\dot\Phi_4)^2\notag\\
&+\frac{1}{2}C_{34}(\dot\Phi_3-\dot\Phi_4)^2+\frac{1}{2}C_{03}\dot\Phi_3^2+\frac{1}{2}C_{04}\dot\Phi_4^2+ \frac{1}{2}C_{45}(\dot\Phi_4-\dot\Phi_5)^2+\frac{1}{2}C_{46}(\dot\Phi_4-\dot\Phi_6)^2+\frac{1}{2}C_{35}(\dot\Phi_3-\dot\Phi_5)^2\notag\\
&+\frac{1}{2}C_{56}(\dot\Phi_5-\dot\Phi_6)^2+\frac{1}{2}C_{05}\dot\Phi_5^2+\frac{1}{2}C_{06}\dot\Phi_6^2 +\frac{1}{2}C_{14}(\dot \Phi_{4}-\dot\Phi_{1})^2 +\frac{1}{2}C_{36}(\dot \Phi_{6}-\dot\Phi_{3})^2\notag\\[1em]
U &= -E_{J1}\cos[2\pi(\Phi_2-\Phi_1)/\Phi_0+2\pi\Phi_{01}/\Phi_0]-E_{Jc}\cos[2\pi(\Phi_4-\Phi_3)/\Phi_0+2\pi\Phi_{0c}/\Phi_0]\notag\\
&-E_{J2}\cos[2\pi(\Phi_6-\Phi_5)/\Phi_0+2\pi\Phi_{02}/\Phi_0],\notag
\end{align}
where $T$ and $U$ are respectively the kinetic and potential energy, $\Phi_{k}$s are node fluxes and $E_{Jk} (\Phi_{ek}) = \sqrt{E_{JSk}^2+E_{JLk}^2+2E_{JSk}E_{JLk}\cos(2\pi\Phi_{ek}/\Phi_0)}$,
$\Phi_{0k} = \tan^{-1}\left[\frac{E_{JSk}-E_{JLk}}{E_{JSk}+E_{JLk}} \tan(\Phi_{ek}/2)\right]\label{phi0}, ~k \in \{1,2,c\} $ and $E_{J} = E_{JLk}+E_{JSk}$ with $E_{JLk}$ and $E_{JSk}$ being the Josephson junction energies of the superconducting quantum interference derive (SQUID). Introducing a new set of flux variables as 
\begin{align}
    \Phi_{1p/m} &= \Phi_{2}\pm \Phi_1,~~~~~
    \Phi_{cp/m} = \Phi_{4}\pm \Phi_3,~~~~~
     \Phi_{2p/m} = \Phi_{6}\pm \Phi_5,
\end{align}
and the corresponding conjugate charge variables, $Q_{kp/m} = \partial \mathcal{L}/\partial \dot \Phi_{k p/m},~k\in\{1,2,c\}$, we can construct the Hamiltonian using
\begin{align}
    H = \frac{1}{2}\mathbf{Q}\mathbf{C}^{-1}\mathbf{Q}^{T} -E_{J1}\cos(\phi_{1m}+\phi_{01})-E_{Jc}\cos(\phi_{cm}+\phi_{0c})-E_{J2}\cos(\phi_{2m}+\phi_{02}),
\end{align}
where $\phi_{l} = 2\pi \Phi_{k}/\Phi_{0}$, $\phi_{0k} = 2\pi \Phi_{0k}/\Phi_{0}$, $l\in \{1,2,...6\}$ and $\phi_{1m} = \phi_2-\phi_1$, $\phi_{2m} = \phi_6-\phi_5$, $\phi_{cm} = \phi_4-\phi_3$, and  $\mathbf{Q} =\text{vec}(Q_{1p},Q_{1m},Q_{cp},Q_{cm},Q_{2p},Q_{2m})$ and the capacitance matrix is given by
\begin{equation}
    \mathbf{C} = \frac{1}{4}
\left(
\begin{array}{cccccc}
    C_{1p}& C_{1m}&-C_{1pp} & C_{1pm}& 0& 0 \\
    C_{1m}& C_{1p}+4C_{12}&C_{1mp} & C_{1mm}&0& 0\\
    -C_{1pp}& C_{1mp} & C_{cp}& C_{cm}&-C_{2pp} & C_{2pm} \\
    C_{1pm}&C_{1mm} & C_{cm} & C_{cp}+4C_{34} & C_{2mp} & C_{2mm} \\
    0& 0& -C_{2pp}& C_{2mp}& C_{2p} & C_{2m} \\
   0& 0 & C_{2pm} & C_{2mm} & C_{2m}& C_{2p}+4C_{56},
\end{array}
\right)
\end{equation}
where
\begin{align}
    C_{1p/m} &=  C_{02}  + C_{23} + C_{24} \pm (C_{01} + C_{13}+C_{14}),\\
    C_{cp/m} &=  C_{24}+ C_{04}+ C_{45} + C_{46} +C_{14}\pm( C_{13} + C_{23} + C_{35} +C_{03} +C_{36}),\\
    C_{2p/m} &=  C_{06} + C_{46} + C_{36}\pm (C_{05}+ C_{35} + C_{45}),\\
    C_{1pp} &= C_{13}+C_{23} + C_{24} +C_{14},~~~~~C_{1pm} = C_{13}+C_{23}-C_{24}-C_{14},\\
    C_{1mp} &= C_{13} +C_{14}-C_{23}-C_{24},~~~~~C_{1mm} = C_{14}+C_{23}-C_{24}-C_{13},\\
    C_{2pp} &= C_{35} + C_{36} + C_{45} + C_{46}, ~~~~~ C_{2pm} = C_{35}+C_{45}-C_{46}-C_{36} ,\\
    C_{2mp} &= C_{35}+C_{36}-C_{45}-C_{46},~~~~~ C_{2mm} = C_{36}+C_{45} -C_{35}-C_{46}.
\end{align}

Note that the qubits and the coupler modes are represented by the charge variables $Q_{1m}$, $Q_{2m}$, and $Q_{cm}$. The modes represented by $Q_{1p}$, $Q_{2p}$, and $Q_{cp}$ are 
``free particles'' rather than harmonic oscillators because their spring constants vanish (no inductances associated with these modes). (Mathematically the Lagrange equation corresponding to these modes is $d (\partial \mathcal{L}/\partial \dot \Phi_{jp})/dt=0$, indicating that the modes are constants of motion.) 
These modes simply correspond to a uniform net charge distributed on the qubits' or coupler's pads \cite{Devoret07}, so we ignore them altogether in the final Hamiltonian.

Keeping the qubits' and coupler's modes only, the Hamiltonian can be written as
\begin{align}\label{H_floating}
    H  =4E_{C1}\hat n_1^2+4E_{C2}\hat n_2^2 + 4E_{Cc}\hat n_c^2 
    + 4E_{12}\hat n_1 \hat n_2 + 4E_{1c} \hat n_1 \hat n_c + 4E_{2c}\hat n_2 \hat n_c
    -\sum_{k=1,2,c}E_{Jk}\cos(\hat\phi_{km}+\phi_{0k}),
\end{align}
where  $\hat n_{k} = Q_{k}/2e$, $E_{Ck}$ ($k\in \{1,2,c\}$), $E_{12},E_{1c},E_{2c}$ are Cooper pair number operators, charging energies, and coupling energies, respectively, and are given by
\begin{align}
    E_{C1} &=E_{C2}\approx \frac{e^2}{2C_{q}+C_g},\\
    E_{Cc} & \approx  \frac{e^2}{2C_{c}+C_{gc}},\\
    E_{12} & \approx -\frac{e^2}{C_{gc}(2 C_{q} + C_g)^2 (2 C_{c} + C_{gc})} 
    \left\{[(C_{23}+C_{24})(C_{45}+C_{35})+C_{24}C_{35}]C_c + (C_{23}C_{35}+C_{45}C_{24})C_{gc}\right\},\label{E12}\\
    E_{1c} &\approx -\frac{e^2(C_{23} - C_{ 24})}{(2 C_{q} + C_g) (2 C_{ c} + C_{gc})},\label{E1c}\\
     E_{2c} &\approx -\frac{e^2(C_{45} - C_{ 35})}{(2 C_{q} + C_g) (2 C_{c} + C_{gc})},
  \label{E2c}
\end{align}
where we have assumed for simplicity the capacitances of the qubits to be the same: $C_{q}= C_{12}= C_{56}$ is the qubit capacitance parallel to the Josephson junction (JJ) and includes the JJ capacitance, $C_{g} = C_{02}= C_{05} = C_{01} = C_{06}$ is the capacitance from capacitor pads to the ground. We assume that $C_{13} = C_{46} =0$ and the capacitances from coupler's pads to the ground are the same $C_{gc}= C_{03}= C_{04}$. In general, the capacitance to the ground is the largest capacitance, $C_{g}>C_{gc}>\{C_{q},C_{c}\}$. Although we cannot say much about the magnitude of the coupling energies $E_{12}, E_{1c}$, and $E_{2c}$, we can easily see from Eqs. \eqref{E12} the sign of $E_{12}$ always negative and the signs of the coupling energies $E_{jc}$ depend on the magnitude of coupling capacitances $C_{23},C_{45},C_{24}$, and $C_{35}$ per Eqs.~\eqref{E1c} and \eqref{E2c}.

The Hamiltonian \eqref{H_floating} can be expressed in the harmonic oscillator basis by introducing annihilation and creation operators for a harmonic oscillator as
\begin{align}
&\hat n_{k} = in_{k}^{\rm zpf}\left(a_k^{\dag}-a_k\right),~~~~\hat \varphi_{k} = \phi_{k}^{\rm zpf}\left(a_k^{\dag}+a_k\right), ~~k\in \{1,2,c\}.
\end{align}
The zero-point fluctuations(zpf) are given by
\begin{align}
    n_{k}^{\rm zpf} &= \frac{1}{\sqrt{2}}\left(\frac{E_{Jk}}{8E_{Ck}}\right)^{\frac{1}{4}},~~~~
    \varphi_{k}^{\rm zpf} =\frac{1}{\sqrt{2}}\left(\frac{8E_{Ck}}{E_{Jk}}\right)^{\frac{1}{4}},
\end{align}
where $[a_k,a^{\dag}_k]=1$. Substituting these in Eq.~\eqref{H_floating} and expanding the cosine terms up to the fourth order, we get    
\begin{align}H & = \sum_{k=1}^{2,c}\left[\omega_{k}+\frac{E_{Ck}}{2}(1+\frac{\xi_k}{4}) -\frac{E_{Ck}}{2}(1+\frac{9\xi_k}{16})a_k^{\dag}a_k\right]a_k^{\dag}a_k\notag\\
    &+ \sum_{j=1}^2 g_{jc}\left(a_{j}a_{c}^{\dag}+a_{j}^{\dag} a_{c}-a_{j}a_{c}-a_{j}^{\dag} a_{c}^{\dag}\right) + g_{12}\left(a_{1}a_{2}^{\dag}+a_{1}^{\dag} a_{2}-a_{1}a_{2}-a_{1}^{\dag} a_{2}^{\dag}\right),
\end{align}
where the frequencies of the qubits and the tunable coupler are:
\begin{align}
    \omega_{k} & = \sqrt{8E_{Jk}E_{Ck}}-E_{Ck}(1+\xi_k/4),~~~\xi_{k} = \sqrt{2E_{Ck}/E_{Jk}}, ~k \in \{1,2,c\}
\end{align}
and the coupling strengths are
\begin{align}
    g_{jc} &= \frac{E_{jc}}{\sqrt{2}}\left(\frac{E_{Jj}}{ E_{Cj}} \frac{E_{Jc}}{E_{Cc}}\right)^{\frac{1}{4}}[1-\frac{1}{8}(\xi_{c}+\xi_{j})],~j\in\{1,2\},\label{g12_appedix}\\
    g_{12} &= \frac{E_{12}}{\sqrt{2}}\left(\frac{E_{J1}}{ E_{C1}} \frac{E_{J2}}{E_{C2}}\right)^{\frac{1}{4}}[1-\frac{1}{8}(\xi_{1}+\xi_{2})].\label{virtual2}
\end{align}
It is worth noting the signs of the coupling strengths $g_{jc}$ and $g_{12}$ are determined by the coupling energies $E_{jc}$ and $E_{12}$, respectively. The floating tunable coupler presented in this work relies on engineering these coupling energies to control the polarity of $g_{jc}$ so that the virtual interaction mediated by the coupler can offset the direct fixed coupling $g_{12}$ at a certain coupler frequency.

\section{Derivation of Hamiltonian of grounded qubits coupled via a floating tunable coupler}\label{ground}

Here we derive the coupling energies for grounded qubits coupled via floating coupler shown in the main text. Introducing node fluxes $\Phi_l~(l=1,...,6)$ and external flux biases $\Phi_{ek}~ (k=1,2,c)$ through each SQUID loop, the Lagrangian of the circuit shown in Fig.~\ref{fig:grounded}(a) can be written as
\begin{align}
    \mathcal{L} & = T-U\\
    T &= \frac{1}{2}C_{1}\dot\Phi_1^2 +\frac{1}{2}C_{2}\dot\Phi_4^2+\frac{1}{2}C_{ c}(\dot\Phi_3-\dot \Phi_2)^2+\frac{1}{2}C_{gc}(\dot\Phi_2^2 +\dot\Phi_3^2)\notag\\
    &+\frac{1}{2}C_{12}(\dot\Phi_2-\dot\Phi_1)^2+\frac{1}{2}C_{13}(\dot\Phi_3-\dot\Phi_1)^2
    +\frac{1}{2}C_{34}(\dot\Phi_4-\dot\Phi_3)^2+\frac{1}{2}C_{24}(\dot\Phi_4-\dot\Phi_2)^2\notag\\
    &U=-E_{J1}\cos(2\pi\Phi_1/\Phi_0 +2\pi\Phi_{e1}/\Phi_0)-E_{J2}\cos(2\pi\Phi_4/\Phi_0 +2\pi\Phi_{e2}/\Phi_0)-E_{J1}\cos[2\pi(\Phi_3-\Phi_2)/\Phi_0 +2\pi\Phi_{ec}/\Phi_0],
\end{align}
where $E_{Jk}$ are junction energies and we have assumed that the capacitances from the coupler's pads to the ground are the same, $C_{gc}= C_{02}=C_{03}$. Introducing new flux variables $\Phi_{p/m} = \Phi_{3}\pm \Phi_2$ and defining conjugate variables $Q_1 = \partial \mathcal{L}/\partial \dot\Phi_1$, $Q_2 = \partial \mathcal{L}/\partial \dot\Phi_4$, $Q_{cp} = \partial \mathcal{L}/\partial \dot\Phi_{cp}$, and $Q_{cm} = \partial \mathcal{L}/\partial \dot\Phi_{cm}$, the corresponding Hamiltonian is given by
\begin{align}
    H= \frac{1}{2}\mathbf{Q}\mathbf{C}^{-1}\mathbf{Q}^T+U
\end{align}
where the capacitance matrix is 
\begin{equation}
    \mathbf{C} = 
\left(
\begin{array}{cccc}
   C_{\Sigma 1} & -\frac{1}{2}(C_{12}+C_{13}) & \frac{1}{2}(C_{12}-C_{13})& 0\\
   -\frac{1}{2}(C_{12}+C_{13})& \frac{1}{4}C_{\Sigma p}& \frac{1}{4}(C_{34}-C_{24}+C_{13}-C_{12})& -\frac{1}{2}(C_{34}+C_{24})\\
   \frac{1}{2}(C_{12}-C_{13}) &\frac{1}{4}(C_{34}-C_{24}+C_{13}-C_{12}) &\frac{1}{4}(C_{\Sigma p}+4C_{c}) & -\frac{1}{2}(C_{34}-C_{24})\\
   0 & -\frac{1}{2}(C_{34}+C_{24}) & -\frac{1}{2}(C_{34}-C_{24}) & C_{\Sigma 2}
\end{array}
\right),
\end{equation}
where $C_{\Sigma 1} = C_1+C_{12}+C_{13}$, $C_{\Sigma 2} = C_2+C_{34}+C_{24}$, and $C_{\Sigma p}= C_{12}+C_{13}+C_{34}+C_{24}+2C_{gc}$. Keeping the modes represented by $Q_1$, $Q_2$ and $Q_{cm}$, the Hamiltonian can put in the form
\begin{align}\label{H_grounded}
    H  =4E_{C1}\hat n_1^2+4E_{C2}\hat n_2^2 + 4E_{Cc}\hat n_c^2 
    + 4E_{12}\hat n_1 \hat n_2 + 4E_{1c} \hat n_1 \hat n_c + 4E_{2c}\hat n_2 \hat n_c
    -\sum_{k=1,2,c}E_{Jk}\cos(\hat\phi_{km}+\phi_{0k}),
\end{align}
where  $\hat n_{k} = Q_{k}/2e$ are the number of Cooper-pair operators, $E_{Ck}$ ($k\in \{1,2,c\}$) are the charging energies; $E_{12}$ and $E_{jc}$ are the qubit-qubit and qubit-coupler coupling energies, respectively. The coupling and charging energies are 
\begin{align}
    E_{12} = \frac{e^2}{4C_{\rm tot}}[C_{12}C_{34}(C_{13}+C_{24}+C_c)+C_{13}C_{24}(C_{12}+C_{34}+C_{c})+
    (C_c+C_{gc})(C_{12}C_{24}+C_{13}C_{34})],
\end{align}
\begin{align}
    E_{1c} = -\frac{e^2}{C_{\rm tot}}[C_2(C_{12}C_{34}-C_{13}C_{24})+C_{gc}C_{\Sigma 2}(C_{12}-C_{13})],
\end{align}
\begin{align}
    E_{2c} = \frac{e^2}{C_{\rm tot}}[C_1(C_{12}C_{34}-C_{13}C_{24})+C_{gc}C_{\Sigma 1}(C_{34}-C_{24})],
\end{align}
\begin{align}
    E_{C1} &\approx  \frac{e^2}{2C_{\rm tot}}\left\{C_2C_{12}C_{34}+C_c C_{gc}(C_2+C_{34})+ [(C_{12}+C_{gc})(C_2+C_{34})+C_2C_{34}](C_c+C_{gc})\right\},
\end{align}
\begin{align}
    E_{C2} \approx \frac{e^2}{2C_{\rm tot}}\left\{C_1C_{12}C_{34} +C_c C_{gc}(C_1+C_{12})+[(C_{34}+C_{gc})(C_1+C_{12})+C_1 C_{12}](C_c+C_{gc})\right\},
\end{align}
\begin{align}
    E_{Cc}\approx \frac{e^2}{C_{\rm tot}}\left[C_{1}C_{12}(C_2+C_{34})+C_{2}C_{34}(C_1+C_{12})+2(C_1+C_{12})(C_2+C_{34})C_{gc}\right],
\end{align}

where 
\begin{align}
    C_{\rm tot} \approx C_{1}C_{2}C_{12}C_{34}+C_{gc}(C_1+C_{12})(C_2+C_{34})\left(2C_{c}+C_{gc}\right) +(C_c+C_{gc})\left[C_{1}C_{12}(C_2+C_{34})+C_2C_{34}(C_1+C_{12})\right].
\end{align}
Expressing the Cooper pair and reduced flux operators in terms of the harmonic oscillator basis, the coupling rates will have the same form as Eqs. \eqref{g12_appedix} and \eqref{virtual2}.
\end{widetext}

%
\end{document}